# Intrinsic defects and mid-gap states in quasi-one-dimensional Indium Telluride

Meryem Bouaziz[1], Aymen Mahmoudi[1], Geoffroy Kremer[1,2], Julien Chaste[1], Cesar Gonzalez[3,4], Yannick J. Dappe[5], François Bertran[6], Patrick Le Fèvre[6], Marco Pala[1], Fabrice Oehler[1], Jean-Christophe Girard[1], Abdelkarim Ouerghi[1*]

[1]Université Paris-Saclay, CNRS, Centre de Nanosciences et de Nanotechnologies, 91120, Palaiseau, Paris, France
[2]Institut Jean Lamour, UMR 7198, CNRS-Université de Lorraine, Campus ARTEM, 2 allée André Guinier, BP 50840, 54011 Nancy, France
[3]Departamento de Física de Materiales, Universidad Complutense de Madrid, 28040 Madrid, Spain
[4]Instituto de Magnetismo Aplicado UCM-ADIF, E-28232 Las Rozas de Madrid, Spain
[5]SPEC, CEA, CNRS, Université Paris-Saclay, CEA Saclay, Gif-sur-Yvette Cedex 91191, France
[6]Synchrotron-SOLEIL, Saint-Aubin, BP48, Paris, F91192, Gif sur Yvette, France

Recently, intriguing physical properties have been unraveled in anisotropic semiconductors, in which the in-plane electronic band structure anisotropy often originates from the low crystallographic symmetry. The atomic chain is the ultimate limit in material downscaling for electronics, a frontier for establishing an entirely new field of one-dimensional quantum materials. Electronic and structural properties of chain-like InTe are essential for better understanding of device applications such as thermoelectrics. Here, we use scanning tunneling microscopy/spectroscopy (STM/STS) measurements and density functional theory (DFT) calculations to directly image the in-plane structural anisotropy in tetragonal Indium Telluride (InTe). As results, we report the direct observation of one-dimensional $In^{1+}$ chains in InTe. We demonstrate that InTe exhibits a band gap of about $0.40 \pm 0.02$ eV located at the M point of the Brillouin zone. Additionally, line defects are observed in our sample, were attributed to $In^{1+}$ chain vacancy along the c-axis, a general feature in many other TlSe-like compounds. Our STS and DFT results prove that the presence of $In^{1+}$ induces localized gap state, located near the valence band maximum (VBM). This acceptor state is responsible for the high intrinsic p-type doping of InTe that we also confirm using angle-resolved photoemission spectroscopy.

## I. INTRODUCTION

In the vast family of two dimensional (2D) (vdW) materials[1], many iconic compounds such as graphene, boron nitride or transition metal dichalcogenide ($MX_2$, M being a transition metal and X a chalcogen element[2–4]) display an hexagonal crystal symmetry. Consequently, their out-of-plane (z direction) properties (electronic, mechanic, thermal, …) are usually very different from their in-plane counterpart (xy directions), but the material is essentially isotropic in the xy plane due to the six-fold crystal symmetry[5]. In addition to those 2D in-plane isotropic compounds, quasi-one-dimensional (quasi-1D) crystals are a class of anisotropic materials wherein atoms are arranged within the xy plane to form 1D-like structures, most often covalent chains, which extend along a particular lattice direction[6,7]. Some examples of these quasi-1D crystals include, black phosphorus (BP), SnSe, GaTe, InTe and $ReS_2$[8–12], which have attracted considerable attentions owing to their in-plane anisotropic physical properties, including high electronic conductivity, thermal conductivity, and exciton recombination occurring only for a given lattice direction[10]. Such particular properties have prompted unconventional applications, such as polarization-sensitive photodetectors and thermoelectric devices[12,13]. As a consequence of their quasi-1D crystal structure, the electronic band structure of these materials typically combines the 2D-like thickness-dependent character along the z direction, with inequivalent electronic dispersions along the x- and y-directions in the plane, bridging the gap between 2D and 1D materials such as nanowires, nanotubes, and other systems with a high geometrical aspect ratio[7].

Mono-telluride compounds such as InTe and GaTe typically show a quasi-1D crystal structure, with weak out-of-plane bonding and anisotropic in-plane character[11]. However their environmental stability under ambient conditions is of concern, since other tellurium-based 2D materials such as $MoTe_2$ and $WTe_2$ are known to rapidly suffer from oxygen or moisture exposure[14,15]. As a consequence, the surface investigation of pristine InTe requires specific conditions, which we propose here to conduct using scanning tunneling microscopy (STM) operating in ultra-high vacuum (UHV) conditions. STM has long been used to image the electronic structure of individual point defects in conductors, semiconductors, and ultrathin films[16]. Specific defects in InTe such as In-vacancies are known to form in the bulk and to affect thermoelectric properties[17,18]. In this context, the experimental surface characterization of the quasi-1D InTe surface and its associated defects is of interest but not yet reported in the literature.

Here, we investigate the surface and electronic properties of InTe using STM and angle-resolved photoemission spectroscopy (ARPES). The crystalline nature of tetragonal InTe is first confirmed by STM, micro-Raman spectroscopy and ARPES, with the expected natural cleavage plane corresponding to the (110) surface[10]. The ARPES data presents a sharp band structure, indicative of the quality of the InTe crystal, but also a substantial p-type character. Further investigations by scanning tunneling spectroscopy (STS) identify specific defective regions, which imposes particular electronic properties on InTe with respect to other defect-free area. Comparison with theoretical band structure and local density of state (LDOS) calculation using density functional theory (DFT) confirms the existence of in-gap states close to the valence band maximum (VBM), at the origin of the observed p-type character of InTe.

## II. EXPERIMENTAL DETAILS

**Photoemission spectroscopy.** ARPES experiments were performed at the CASSIOPEE beamline of the SOLEIL synchrotron light source. The CASSIOPEE beamline is equipped with a Scienta R4000 hemispherical electron analyzer whose angular acceptance is ±15° (Scienta Wide Angle Lens). The valence band data were calibrated with respect to the Fermi level. High-quality samples from the "2D semiconductors" company were cleaved in UHV at a base pressure better than $1 \times 10^{-10}$ mbar. The Fermi level was determined by fitting the leading edge of the gold crystal at the same photon energies and under the same experimental conditions. The experiment was performed at T = 30 K. The incident photon beam was focused into a 50 μm spot (in diameter) on the sample surface. All ARPES measurements were performed with a linear horizontal polarization.

**Scanning tunnelling microscopy.** STM experiments are performed at 77 K using an LT-STM (Scienta-Omicron). The sample has been prepared in the same conditions compared to ARPES measurements, cleaved at a base pressure better than $1 \times 10^{-10}$ mbar and transferred immediately in the pre-cooled STM-head. STM images are acquired in the constant current mode.

**μ-Raman measurements.** The μ-Raman measurements were conducted at room temperature, using a commercial confocal Horiba micro-Raman microscope with a ×100 objective and a 532 nm laser excitation. The laser beam was focused onto a small spot having a diameter of ~1 μm on the sample.

## III. RESULTS AND DISCUSSION

Bulk InTe is a tetragonal semiconductor which crystallizes in the TlSe structure under ambient conditions[19]. It is best described by the tetragonal space group *I4/mcm*, with the unit cell (Figure 1(a), black line), being a right-angled cuboid with parameters a=8.444Å in the (100) and (010) directions while a≠c=7.136Å along the [001] axis[17]. As grown bulk InTe typically exhibits a p-type character[10,20]. While Te atoms solely occupy one crystallographic site (8*h*), In atoms are distributed equally over two independent crystallographic sites, named hereafter $In^{1+}$ and $In^{3+}$, respectively. The $In^{3+}$ ions are tetrahedrally coordinated with Te atoms forming covalent infinite edge-sharing tetrahedra along the c direction. Perpendicularly, along the [100] and [010] directions, those infinite $(InTe_2)^-$ covalent chains are weakly bonded to each other and intercalated by $In^{1+}$ ions. These $In^{1+}$ ions are actually surrounded by eight Te atoms in a tetragonal antiprismatic coordination but show large atomic displacement values along the c direction, moving in a tunnel-like configuration between the covalently bound $(InTe_2)^-$ chains[10]. Due to the weak interactions between $(InTe_2)^-$ chains, the crystal tends to cleave along the (110) planes[20] (Figure 1(a), light brown), cutting through the mobile $In^{1+}$ positions. The exposed (110) surface shows a rectangular symmetry with parameter a√2 and c (Figure 1(b)). While the exposed $In^{1+}$ atoms form a regular lattice of smaller parameter a√2/2 and c/2, the underlying Te atoms are positioned so that adjacent $(InTe_2)^-$ along [-110] chains are shifted by c/2 along [001] (see Figure 1(a) and (b), (220) light green plane), which requires boxing the entire surface cell. Representative Raman spectra acquired in 3 different zones (room temperature, 532nm laser) from a freshly cleaved commercial InTe sample (2D Semiconductors) are shown in Figure 1(c). All spectra expose identical Raman features with the same magnitude. The $E_g$ (48 cm⁻¹, 138 cm⁻¹), $B_{1g}$ (86 cm⁻¹) and $A_{1g}$ (126 cm⁻¹) single modes are observed with frequencies that are in agreement with literature on ambient tetragonal InTe[21,19], confirming the expected crystal structure and quality of our sample.

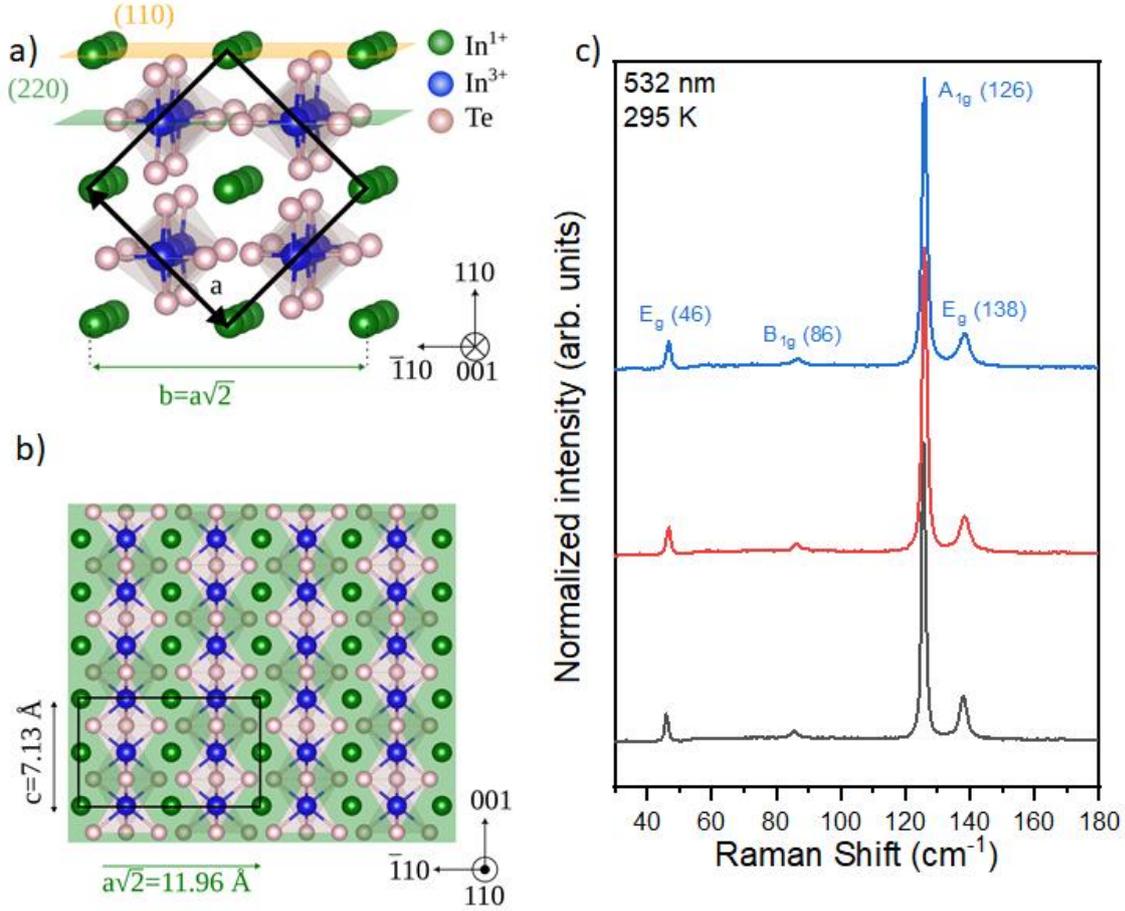

**Figure 1:** Crystallographic and µ-Raman analysis of InTe: (a) Crystal structure of tetragonal InTe. In[1+], In[3+] and Te atoms are represented in blue, purple and orange spheres, respectively. The preferred 110 cleaving plane is indicated in light brown. The unit cell in marked by black line. (b) Theoretical InTe (110) surface, exposing the In[1+] atoms and the corresponding surface cell parameter. (c) Room temperature µ-Raman spectra acquired from different zone from a cleaved InTe (110) surface, showing the 3 expected Raman lines: $Eg$ at 48 cm$^{-1}$ and 138 cm$^{-1}$ and the $A_{1g}$ at 126 cm$^{-1}$.

To resolve the above-described electronic properties of InTe, we have performed low temperature ARPES measurements (30 K). In Figure 2(a), we show the projected first Brillouin zone (BZ) of InTe for the (110) surface. It exposes two inequivalent directions in reciprocal space which are perpendicular: $\overline{MXM}$ which is aligned with real space direction [-110] (i.e. across the (InTe$_2$)$^-$ chains), and $\overline{MZM}$ along [001] (i.e. along the (InTe$_2$)$^-$ chains), as expected from the anisotropic real space symmetry (Figure 1(a, b)). In Figure 2(b) we present the isoenergetic contours obtained at M point (BE= -0.5 eV). The elliptical valley image is the result of anisotropy electronic properties of InTe. In Figure 2(c and d), we report the experimental ARPES dispersion of the (110) InTe surface acquired along the $\overline{MXM}$ direction, using a photon energy of 50 and 118 eV (surface and bulk sensitivity). The top most part of the valence band is characterized by hole-like bands centered at the *M* point. The valence band maxima (VBM) at the M point of the Brillouin zone is mostly formed by p and s orbital of Te and In respectively, while at the X point the band is mostly composed by p orbitals of Te. We observe that the valence band maximum (VBM) is located at the Fermi level for the two-photon energy. This confirms the high p-type character of our InTe (110) bulk. Comparison with DFT calculations using the HSE06 functional (Figure 2(e)) shows a good agreement. Considering the electronic bandgap of InTe calculated by HSE, the VBM position indicates a p-type doping of InTe. This result confirms the strong p-type character of our InTe in agreement with literature results on bulk InTe[10].

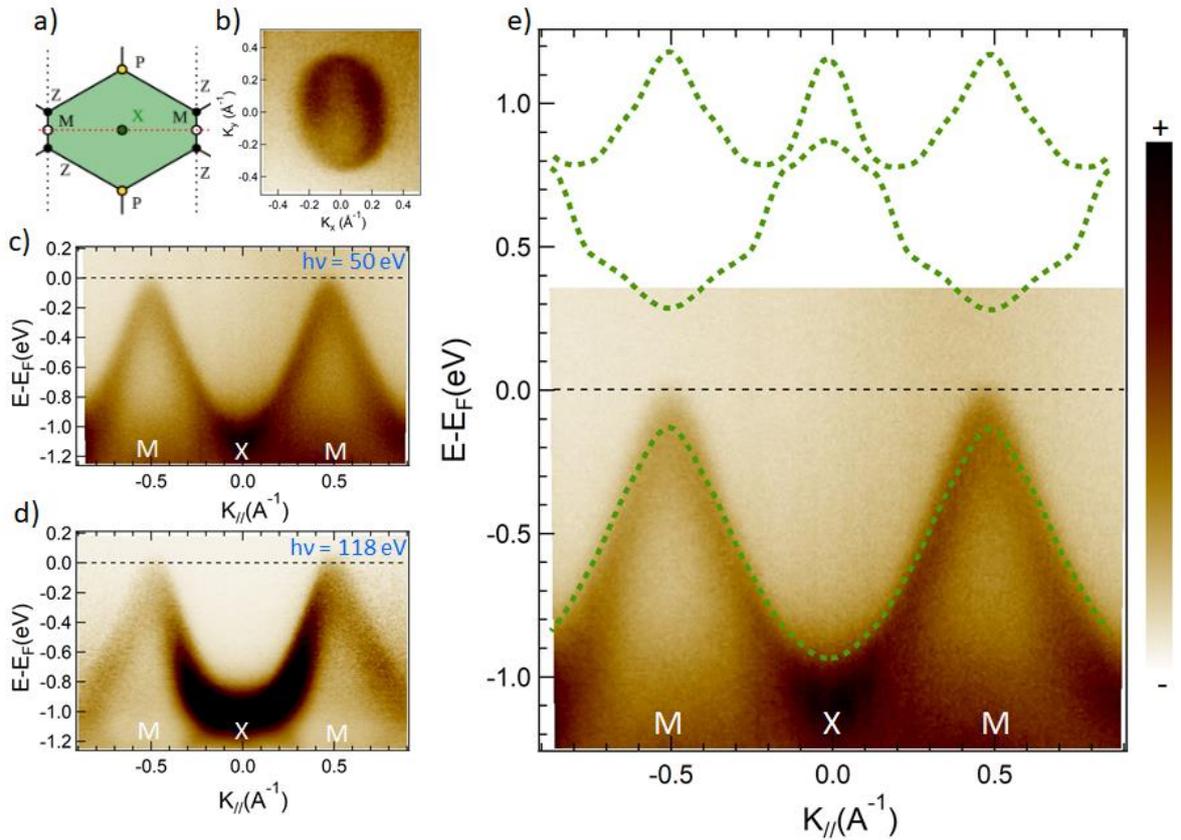

**Figure 2:** (a) Projected Brillouin zone of InTe along the 110 axis: b) Isoenergic contour around M point at -0.5 eV showing the 1D character of InTe (hʋ= 50 eV), (c and d) Experimental ARPES dispersion of InTe along the MXM direction obtained with a photon energy of hʋ= 50 eV and 118 eV, respectively. (d) ARPES data (hʋ= 50 eV) superposed with theoretical DFT results computed with the HSE06 functional. The DFT bands are shifted to account for the Fermi level position for the ARPES image.

In order to determine the physical origin of the observed strong p-type doping, we carried out a combined STM/STS study of the InTe (110) surface[22,7]. Figure 3(a) shows a large-scale topographic STM image of the UHV-cleaved surface of InTe, revealing wide terraces separated by high density step edges and extended line defects aligned exclusively along a single in-plane direction, close to the vertical axis in Figure 3(a). In zoomed-in STM images, Figure 3(b) and (c), we notice the ordered flat regions are better resolved and show finer features aligned along the same direction than the step edges. The corresponding Fourier transform (FFT); shown in Figure 3(d) reveals a clear rectangular symmetry, with the spacing along the vertical direction being close to that of c-axis (8.0 Å) and the horizontal matching that of a√2 = (11.9 Å). Comparison with the expected InTe (110) surface (Figure 1(b)), confirms the surface orientation and cleavage plane, and determine the long-axis of the step edges (large scale, Figure 3(a),) and linear atomic features (small scale, Figure 3(b-c)) to be the crystal c-axis. The absence of any other feature indicates that our InTe crystal is fully monocrystalline, without any rotational disorder, twinning domains, pits or antiphase domain boundaries. The STM characterization of the InTe (110) surface strongly reflects the expected anisotropy of the InTe (110) and allows the direct imaging of 1D-like electronic features along the (in-plane) c-axis[10,19]. Figure 3(c) clearly shows that the InTe layer is incomplete, i.e. a $In^{1+}$ and $(InTe_2)^-$ ions are missing, which leads to the appearance of dark lines corresponding to holes. This suggests that single-atom vacancy defects on InTe can be formed. The density of $In^{1+}$ vacancies is estimated about 20-40 % of the area.

To investigate the electronic band gap, complementary STS measurement at 77K in defect-free (red spectrum) and defective area (blue spectrum), coupled to DFT calculations of local density of states (LDOS)[23] are presented Figure 3(e). The experimental differential conductance (dI/dV) spectroscopy versus bias voltage spectra, is compared to the LDOS calculated by DFT. The theoretical LDOS of the InTe is very sharp at the top of the valence band and near the conduction band minimum as well as for the experimental STS. The conduction band minimum (CBM) is located at $0.28 \pm 0.02$ eV above the Fermi level, for both experimental spectra as well as for the calculated LDOS. However, a difference on the valence band side is noticed. The valence band maximum (VBM) is located at $0.12 \pm 0.02$ eV below the Fermi level (i.e., zero bias on the dI/dV spectra), in the defect-free area, while it is located near $E_F$ for the spectrum measured on a defect. By comparing the two STS spectra (red and blue tank), we can see a difference on the VBM region. This difference can be associated with localized states in the gap of the sample near the VBM. The intrinsic electron quasiparticle bandgap is estimated about E= $0.40 \pm 0.02$ eV. This value is of the same range as the one obtained with the DFT (~0.41 eV). The uncertainty in bandgap is the result of the lateral band edge variations. The Fermi energy, corresponding to the zero bias in the dI/dV spectrum, is positioned near the VBM. As such, it suggests the presence of intrinsic charged defects in our specimens and a p-type doping, which we could relate to lattice vacancies or antisites, responsible for p-doping in other 2D materials[24]. Still, the p-type character of our InTe (110) surface observed by STS agrees with our ARPES measurements.

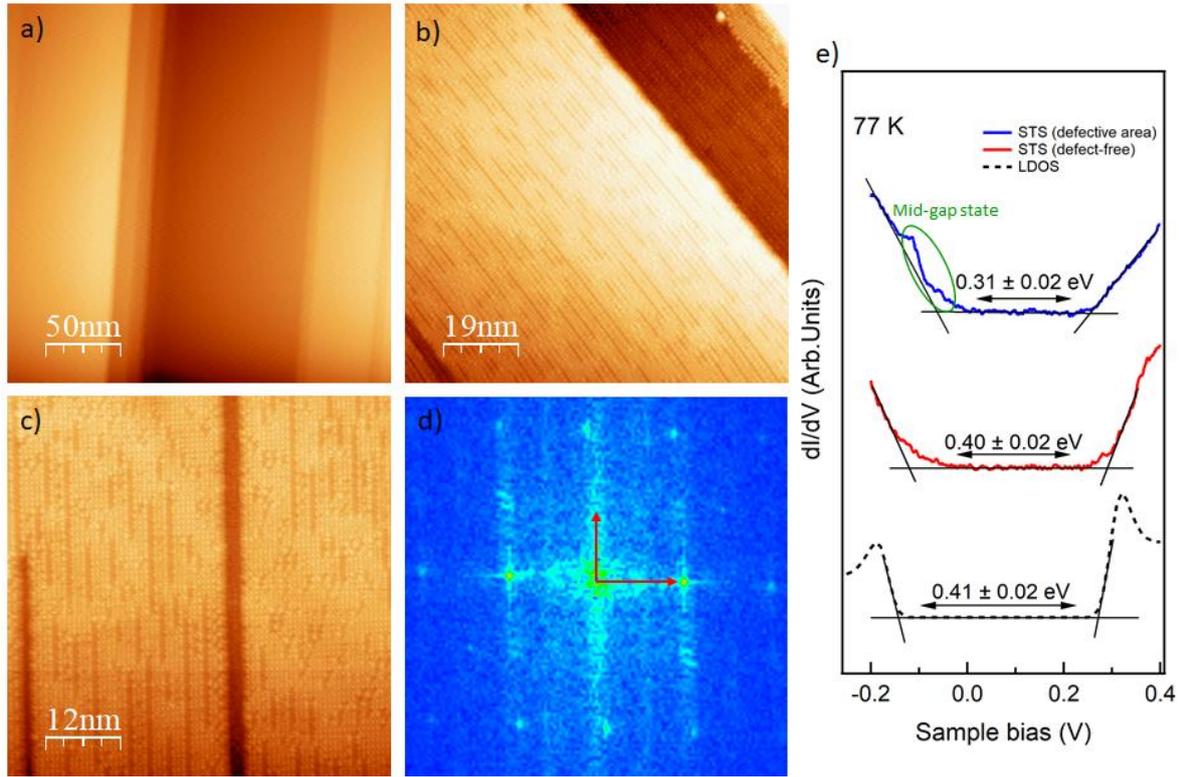

**Figure 3:** Structural and electronic properties of InTe: (a) Large-scale STM image at 77 K of the UHV-cleaved surface of InTe, revealing the oriented step edges on this surface ($V_{bias}$ = -0.5eV, $I_{tunneling}$ = 100 pA). (b and c) Zoomed-in STM images, obtained with the same parameters, showing more linear features. (d) The corresponding fast Fourier transform (FFT) indicating the a√2 x c periodicity of the InTe (110) surface. (e) The differential conductance (dI/dV) spectrum as a function of the bias voltage (red and blue curves) showing that the energy bandgap is ~ 0.40 ± 0.02 eV. Total and partial electronic densities of states (DOS/PDOS) near the top of the valence band and the bottom of the conduction band of an InTe (dashed black curve).

In Figure 4(a), we present a STM image of a particular area (defect free). We measure a spatial periodicity of a√2/2 x c/2 in the perpendicular directions, which is consistent with the lattice parameters of the structure (Figure 1 (b)). From the structural model, we attributed that the imaged bright spots to the $In^{1+}$. To verify our hypothesis, we have performed DFT calculations, using the DFT localized-orbital molecular-dynamics code as implemented in Fireball[25–28]. We have optimized a 1x1 unit cell of fully $In^{1+}$ terminated (defect free) InTe and InTe with $In^+$ defects until the forces went below 0.1 eV Å$^{-1}$. A set of 8 x 8 x 1 k-points has been used for both structural optimization and DOS calculations. STM simulations have been performed within a Keldysh-Green function formalism as described in ref.[29] On the other hand, the tip is modelled by a W-pyramid of 5 atoms coupled to four layers of W(100) with a 5x5 periodicity. The WSxM software has been used for image visualization[30]. In Figure 4(b and c), we present two simulated STM images (InTe defect-free and InTe with $In^+$ defects), constructed with parameter coherent with our experimental STM conditions. We observe a good match between the experimental and simulated STM image of defect free InTe (110) (Figure 4(a) and (b)). Based on DFT calculations, we confirm the direct observation of one-dimensional $In^{1+}$ chains in an InTe (figure 4(a)). The superposed crystal lattice from DFT (Figure 4(b) insert) confirms that the observed protrusions localize close to the $In^{1+}$ atomic positions.

For the defective InTe, the calculated STM image (Figure 4(c)) shows the In$^{1+}$ vacancy as depression which is similar to the experimental observation (Figure 3(c)). The topographic signature of these defects is imaged using STM showing the In$^{1+}$ vacancies as depressions on the InTe surface, notably extended vacancy lines along the c crystal axis. If we refer to Figure 3(c), we now interpret the line defect to be In$^{1+}$-defective. Our STM/STS and ARPES result thus demonstrate such In$^{1+}$ vacancies are responsible for the p-type character of InTe and more generally a typical feature of TlSe-like materials[10].

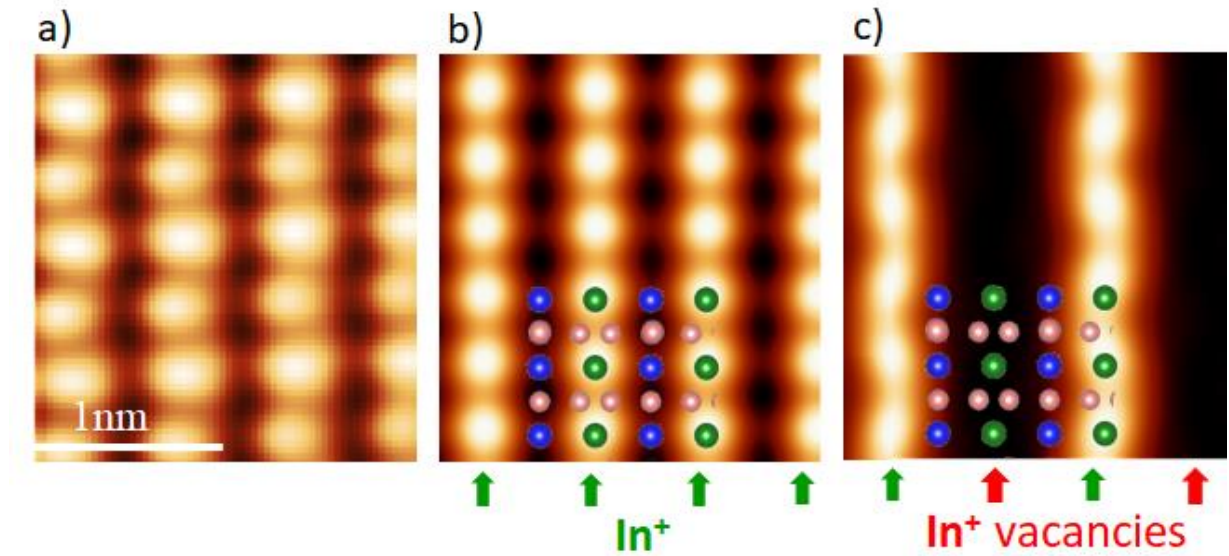

**Figure 4:** Experimental and simulated STM image: (a) Experimental STM image of a fully In$^{1+}$ terminated InTe (110) surface acquired using U = 0.2 V and I = 100 pA. (b and c) Simulated STM image of InTe defect-free and InTe with In$^+$ defects taken at constant height mode with U = −1 V and an STM tip 4.5 Å above the averaged InTe (110) surface. The WSxM software has been used for visualizing the image[30].

We now turn to the detailed investigation of the InTe (110) with In$^{1+}$ defects, in order to study the atomic structure of the observed 1D line features. Figure 5(a) shows an atomically resolved STM image of the InTe (110) surface. The crystal c-axis is vertical and the corresponding chain-like features are now obvious. However, we note that each 1D chain presents a slightly different structure along its axis than its neighbors, alternating short, long protrusions or gap(s). The height profile perpendicular to the chains (Figure 5(a), blue line) is shown in Figure 5(b). The oscillation period is a√2/2 (~ 6 Å) and the depth is about 1 Å. This vertical distance corresponds to the vertical distance between the top In$^{1+}$ atoms and the underlying In$^{3+}$ atoms. The lateral period also matches the expected In$^{1+}$ position, with a spacing of a√2/2 (Figure 1(b)) along the -110 direction. Given that InTe (110) surface is obtained by mechanical cleaving through the In$^{1+}$ (110) plane, we suggest some vacancy in the exposed In$^{1+}$ layer at the top surface. In Figure 5(c) left, we investigate a small regular array characterized by a lateral length of a√2/2, but with a double vertical period (7 Å) equal to cell parameter c. Following our hypothesis, this situation corresponds to a regular reconstruction of the surface with half of In$^{1+}$ atoms missing (Figure 5(c) right). Another variant is presented Figure 5(d) left, in which typical extended 1D defect is shown and that we interpret as In$^{1+}$ vacancies aligned along the c-axis, Figure 5(d) right. This surface model with mobile In$^{1+}$ vacancies is also compatible with the overall picture shown Figure 5(a), in which the spatial distribution of In$^{1+}$ vacancies is more random and create various of combinations of regular array (a$\sqrt{2}$/2 x c, ~ 40% In$^{1+}$ vacancies), vacancy lines or less ordered arrangements.

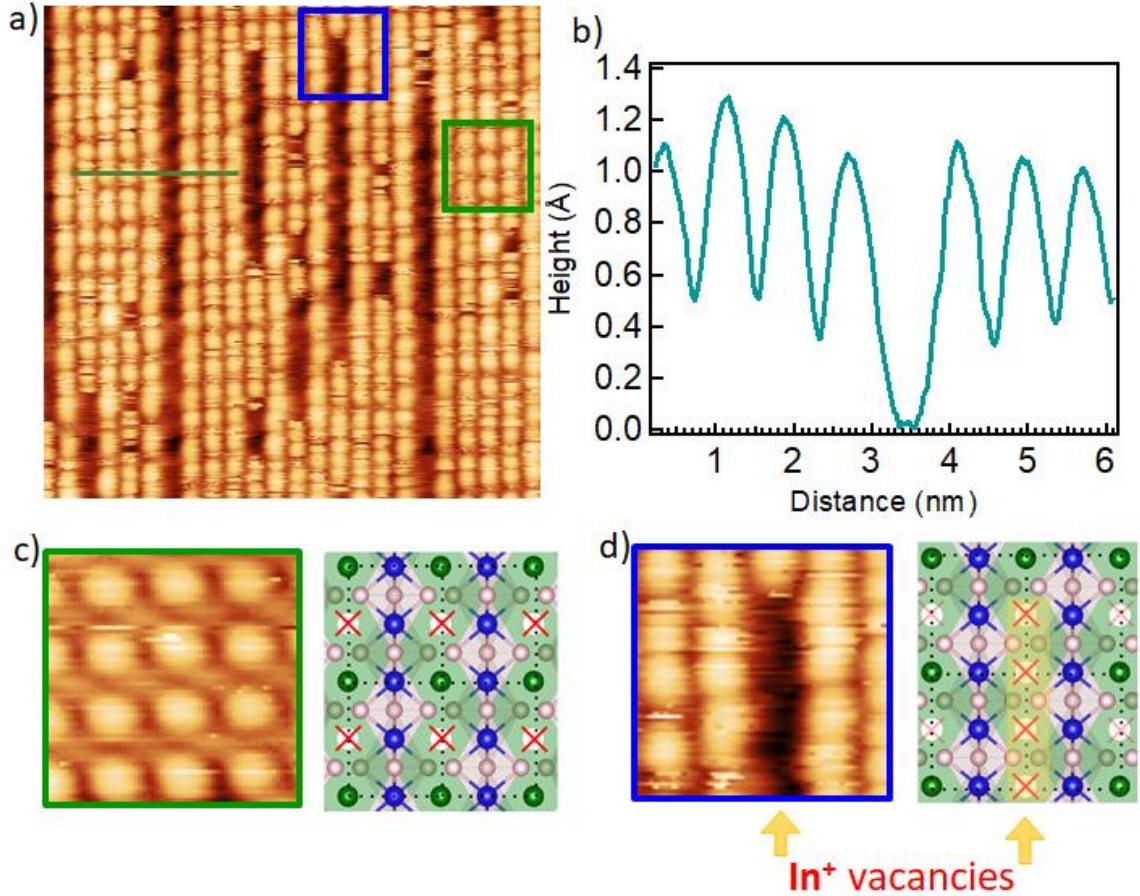

**Figure 5:** Atomistic defects of the InTe (110) surface: (a) Atomically resolved STM topography (26 ×26 Å$^2$, U=−0.5 V, I=30 pA). (b) Line profile across the chain. (c) Regular In$^{1+}$ vacancy arrangement in a rectangular lattice. (d) Linear arrangement of In$^{1+}$ vacancy forming an extended segment along the in-plane c-axis.

## IV. CONCLUSION

In summary, we have combined complementary techniques to investigate the structural and electronic properties of the cleaved surface of quasi-1D InTe. Our combined DFT and STM study confirms the (110) preferential cleaving plane and reveals various surface organization of In$^{1+}$ vacancies, sometimes arranged in regular array but mostly in extended segments along the in-plane c-axis. These In$^{1+}$ vacancies are the dominant defects on the InTe and STS results indicate that they are at the origin of the strong p-type character of InTe, as also observed using ARPES. Further investigation of these In$^{1+}$ vacancies offer insight towards alternative doping strategies, possibly by substitution, to obtain opposite n-type electrical conduction while maintaining the inherent strong anisotropy of InTe.

**Acknowledgments:** We acknowledge the financial support by MagicValley (ANR-18-CE24-0007) and Graskop (ANR-19-CE09-0026), 2D-on-Demand (ANR-20-CE09-0026), and MixDferro (ANR-21-CE09-0029 grants, the French technological network RENATECH. This work is also supported by a public grant overseen by the French National Research Agency (ANR) as part of the "Investissements d'Avenir" program (Labex NanoSaclay, ANR-10-LABX-0035) and by the French technological network RENATECH. C. G. thankfully acknowledges the computer resources at Altamira and the technical support provided by the Physics Institute of Cantabria (IFCA) of the University of Cantabria (UC), project FI-2022-3-0021 and the financial support by the Spanish Science and Innovation Ministry (project PID2021-123112OB-C21).

**Appendix A:**

Figure 6 presents the ARPES dispersion of the (110) InTe surface acquired along the $\overline{MXM}$ direction, using a photon energy of 84 eV.

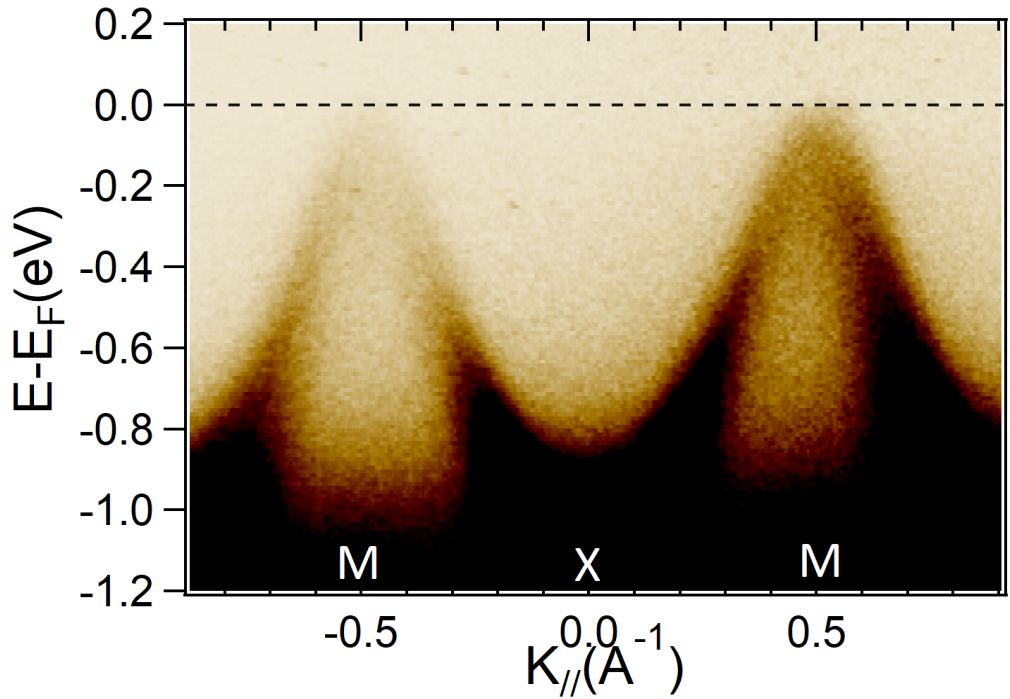

Figure 6. Experimental ARPES dispersion of InTe along the MXM direction obtained with a photon energy of hυ= 84 eV.

The energy distribution curve (EDC) of the valence band, measured with photon energy of 50 and 118 eV, are shown in Figure 7 (green and red lines). On the same graph, the Fermi edge of a clean gold sample measured in the same experimental conditions is presented (blue line). The zero of the binding energy (i.e., the Fermi level) was taken at the leading edge of this gold sample.

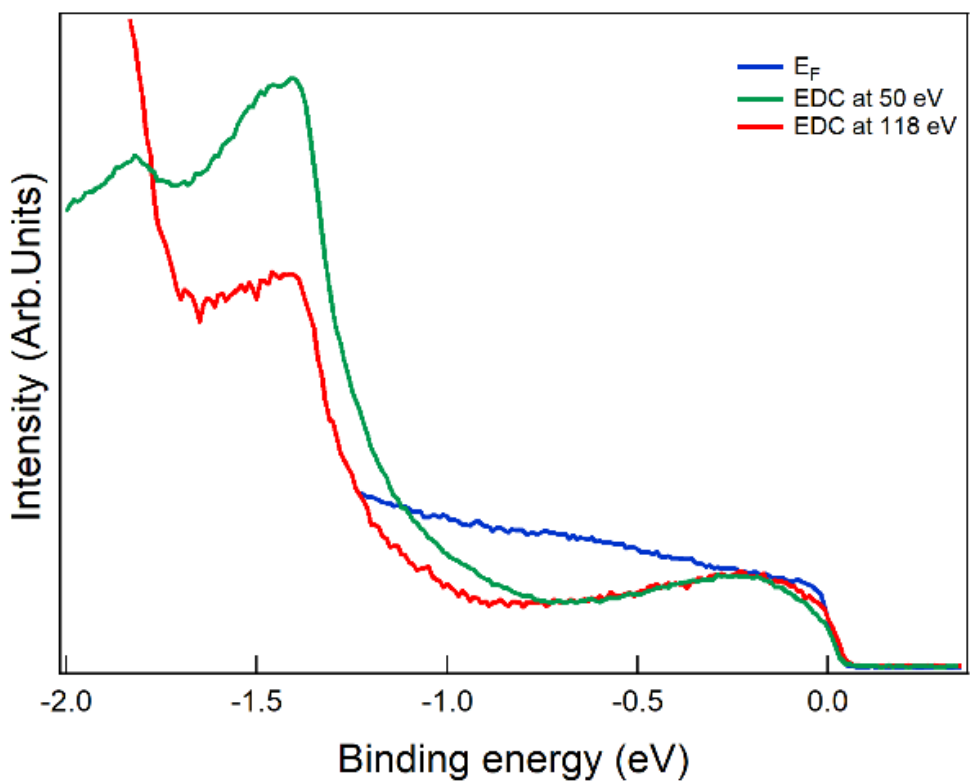

Figure 7. Energy distribution curve (EDC) extracted from ARPES measurements at 50 eV (green line) and 118 eV (red line). The blue curve is the Fermi edge of a clean gold sample measured in the same experimental conditions.